\documentclass[12pt]{article}
\usepackage{amssymb,amsmath,epsfig}

\begin{document}

\title{\bf Cosmic Evolution in Self-Interacting Brans-Dicke Cosmology}

\author{Muhammad SHARIF \thanks{Email address: msharif.math@pu.edu.pk} and Saira
WAHEED
\thanks{smathematics@hotmail.com}\\
Department of Mathematics, University of the Punjab,\\
Quaid-e-Azam Campus, Lahore-54590, Pakistan.}

\date{}

\maketitle
\begin{abstract}
In this paper, we investigate the nature of self-interacting
potential that can support the accelerated expansion of the universe
as indicated by observational data. For this purpose, we consider
the Bianchi type I (BI) universe model in the Brans-Dicke (BD) field
scenario and use a power law for both the scalar field and the scale
factor. It is found that the accelerated expansion of the universe
can be discussed for a positive power law potential with negative
values of the BD parameter. We also explore the evolution of energy
density perturbations. This analysis indicates that the model allows
growing modes for negative values of the BD parameter and $m>0$.
\end{abstract}
{\bf PACs numbers:} 04.50.Kd; 98.80.-k

\section{Introduction}

Many astronomical experiments such as Supernova (Ia) $^{1,2)}$,
Wilkinson Microwave Anisotropy Probe (WMAP) $^{3)}$, Sloan Digital
Sky Survey (SDSS) $^{4)}$, galactic cluster emission of X-rays
$^{5)}$, large scale-structure $^{6)}$ and weak lensing $^{7)}$,
provide evidence of an accelerated expansion of the universe in the
present phase. As the baryonic matter and dark matter do not provide
the total energy density of the universe, it has been suggested that
there is some kind of "missing energy". The Chaplygin gas, phantom,
quintessence and cosmological constant $^{8,9)}$ are some
appropriate candidates of dark energy (missing energy), considered
to be responsible for the expansion of the universe. In order to
conciliate the nucleosynthesis scenario in the radiation-dominated
era, it is argued that the universe was expanding with deceleration
in its early phase.

Quintessence is a dynamical slowly evolving spatially inhomogeneous
component of energy density with a negative pressure $^{9)}$ having
an EoS parameter lying between $0$ and $-1$. However, this model was
proved to be not very successful as it faces problems like fine
tuning and cosmic coincidence $^{10)}$. To solve these problems, a
special form of quintessence field called the \emph{tracker field}
(with a wide range of initial values of $\rho_{\phi}$, such fields
roll down along a common evolutionary track with the present energy
density of the universe $\rho_{n}$ and end up in the observable
universe with $\rho_{\phi}$ comparable to $\rho_{n}$ in the present
epoch) has been constructed $^{11)}$. Different types of potential
have been developed $^{12)}$ to understand the nature of dark energy
using a quintessence model but each of them has its own drawbacks.

Scalar-tensor theories of gravity as modified gravity theories are
proved to be very effective in explaining many outstanding problems
in cosmology $^{13-16)}$ like inflation, the ``early- and late-"
time behaviors of the universe, coincidence problem and cosmic
acceleration. Brans and Dicke $^{17)}$ proposed a scalar-tensor
generalization of general relativity (GR) by introducing a
time-varying gravitational constant (as $G(t)=\frac{1}{\phi(t)}$)
and the direct interaction of a scalar field with a geometry known
as the BD theory of gravity. The weak equivalence principle, Mach's
principle $^{18)}$ and Dirac's large number hypothesis $^{19)}$ are
also accommodated in this theory. Using local gravity tests, it has
been shown $^{20,21)}$ that the generic dimensionless BD parameter
$\omega$ should be very large, i.e., $\omega\geq40,000$, for being
compatible with solar system experiment bounds. This theory goes
over to GR if $\omega \rightarrow \infty$ and the scalar field is
constant $^{22)}$. In general, it has been shown $^{23,24)}$ that
the BD theory is reducible to GR only if the trace of the
energy-momentum tensor $T^{(m)}$ does not vanish, i.e., $T^{(m)}\neq
0$.

Many researchers have explored the cosmic evolution of the universe
in the BD theory. In the GR and scalar tensor theories, much work
has been done using Bianchi models to discuss different stages of
the cosmic evolution $^{25,26)}$. In ref. $27$, it has been pointed
out that some large-angle anomalies are seen in CMB radiations that
violate the statistical isotropy of the observable universe. For a
better description of these anomalies, plane symmetric and
homogeneous but anisotropic universe models play a very significant
role. Moreover, it is pointed out $^{28)}$ that removing a Bianchi
component from WMAP data can explain various large-angle anomalies
yielding an isotropic universe. Thus, the universe may have
accomplished a slight anisotropic geometry in cosmological models
irrespective of inflation. King and Coles $^{29)}$ described the
evolution of magnetized axisymmetric Bianchi I universe in the
presence of vacuum energy. Singh and Rai $^{30)}$ constructed
various BD cosmological models using Bianchi models with a perfect
fluid to explain the evolution of the universe. In our recent study
$^{31)}$, we have investigated some Bianchi type I universe models
using perfect, anisotropic and magnetized anisotropic fluids in
self-interacting BD cosmology. Bermann $^{32)}$ investigated
different cosmological models using a constant deceleration
parameter. Khoury and Weltman $^{33)}$ explored the coupling of a
scalar field with matter of unity order, which is allowed to evolve
cosmologically.

Bertolami and Martins $^{13)}$ obtained accelerated expanding
solutions using quadratic potentials. They also evaluated energy
density perturbations and discussed the growing modes of their
model. By extending the concept of quintessence to flat non
minimally coupled scalar tensor theories, Ritis et al. $^{34)}$
found a family of exact solutions parameterized by the parameter
's'. These solutions are proposed as a class of tracker fields known
as tracking solutions. Sen and Sen $^{35)}$ investigated
accelerating solutions using a potential with a time-dependent mass
squared term in BD cosmology. By introducing the non minimal
coupling term $\psi R\frac{\phi^2}{2}$, different kinds of potential
have been discussed $^{36)}$. Sen and Seshadri $^{37)}$ investigated
the nature of self-interacting potential as well as the evolution of
energy density perturbations in this theory.

In this study, we explore the nature of self-interacting potential
and BD field energy density as dark energy, which can account for
the accelerated expansion of the universe by taking a perfect fluid
and the LRS Bianchi type I (BI) universe into account. This paper is
organized as follows. In the next section, we formulate the field
equations of the self-interacting BD theory for the BI universe. We
discuss solutions by taking a power law ansatz. In \textbf{\S}{3},
we provide the evolution of energy density perturbations by taking a
perturbed form of field equations. We discuss the results in the
last section.

\section{Self-Interacting Potential and Dark Energy}

The action for the BD theory with the self-interacting potential
$V(\phi)$ is given by $^{37)}$
\begin{equation}\label{1}
S=\int d^{4}x\sqrt{-g}[\phi
R-\frac{\omega_{0}}{\phi}\phi^{,\alpha}\phi_{,\alpha}-V(\phi)+L_{m}],\quad\alpha=0,1,2,3.
\end{equation}
Here, $L_{m}$ is the matter part of the Lagrangian and $\omega_{0}$
is the constant BD coupling parameter (where the units $8\pi
G_{0}=c=1$ are imposed). The corresponding field equations derived
from the action (\ref{1}) are
\begin{eqnarray}\label{2}
G_{\mu\nu}&=&\frac{\omega_{0}}{\phi^{2}}[\phi_{,\mu}\phi_{,\nu}
-\frac{1}{2}g_{\mu\nu}\phi_{,\alpha}\phi^{,\alpha}]+\frac{1}{\phi}[\phi_{,\mu;\nu}
-g_{\mu\nu}\Box\phi]+\frac{T_{\mu\nu}}{\phi}+\frac{V(\phi)}{2\phi},\\\label{3}
\Box\phi&=&\frac{T}{3+2\omega_{0}}-\frac{2}{3+2\omega_{0}}[V(\phi)-\phi\frac{V(\phi)_{,\phi}}{2}].
\end{eqnarray}
The terms $T=g^{\mu\nu}T_{\mu\nu},~ \Box=\Delta^{\mu}\Delta_{\mu}$
and $\Delta^{\mu}$ represent the trace of the energy-momentum
tensor, the d'Alembertian operator, and the covariant derivative
respectively.

The homogeneous, anisotropic, and spatially flat BI universe model
is given by $^{38)}$
\begin{equation}\label{4}
ds^{2}=dt^{2}-A^{2}(t)dx^{2}-B^{2}(t)(dy^{2}+dz^{2}),
\end{equation}
where $A$ and $B$ are the scale factors along the transverse
direction $x$ and the two equivalent longitudinal directions $y$ and
$z$, respectively. The energy-momentum tensor for a perfect fluid is
\begin{equation}\label{5}
T_{\mu\nu}=(\rho+P)u_{\mu}u_{\nu}-Pg_{\mu\nu},
\end{equation}
where $\rho$ and $P$ are the energy density and pressure,
respectively, and $u^{\mu}$ is the four-velocity with the
normalization condition $u_{\mu}u^{\nu}=1$. The energy conservation
equation is
\begin{equation}\label{6}
\dot{\rho}+(\frac{\dot{A}}{A}+2\frac{\dot{B}}{B})(\rho+P) =0.
\end{equation}
The corresponding BD field equations become
\begin{eqnarray}\label{7}
&&\frac{2\dot{A}\dot{B}}{AB}+\frac{\dot{B}^{2}}{B^{2}}
=\frac{\rho}{\phi}+\frac{\omega_{0}}{2}
\frac{\dot{\phi}^{2}}{\phi^{2}}-(\frac{\dot{A}}{A}
+2\frac{\dot{B}}{B})\frac{\dot{\phi}}{\phi}+\frac{V(\phi)}{2\phi},\\\label{8}&&
2\frac{\ddot{B}}{B}+\frac{\dot{B}^{2}}{B^{2}}=-\frac{P}{\phi}-\frac{\omega_{0}}{2}
\frac{\dot{\phi}^{2}}{\phi^{2}}-2\frac{\dot{B}}{B}\frac{\dot{\phi}}{\phi}
-\frac{\ddot{\phi}}{\phi}+\frac{V(\phi)}{2\phi},\\\label{9}
&&\frac{\ddot{B}}{B}+\frac{\ddot{A}}{A}+\frac{\dot{A}\dot{B}}{AB}
=-\frac{P}{\phi}-\frac{\omega_{0}}{2}
\frac{\dot{\phi}^{2}}{\phi^{2}}-\frac{\ddot{\phi}}{\phi}-(\frac{\dot{A}}{A}
+\frac{\dot{B}}{B})\frac{\dot{\phi}}{\phi}+\frac{V(\phi)}{2\phi},
\end{eqnarray}
while the Klein-Gordan equation for scalar field is
\begin{equation}\label{10}
\ddot{\phi}+(\frac{\dot{A}}{A}+2\frac{\dot{B}}{B})\dot{\phi}
=\frac{\rho-3P}{2\omega_{0}+3}-\frac{2}{2\omega_{0}+3}(V-\frac{\phi
V_{,\phi}}{2}).
\end{equation}
The average scale factor and mean Hubble parameter respectively turn
out to be
\begin{equation*}
a^3(t)=AB^{2},\quad
H(t)=\frac{1}{3}(\frac{\dot{A}}{A}+2\frac{\dot{B}}{B}),
\end{equation*}
whereas the expansion and shear scalar are
\begin{equation*}
\theta=u^{a}_{;a}=\frac{\dot{A}}{A}+2\frac{\dot{B}}{B},\quad
\sigma=\frac{1}{\sqrt{3}}(\frac{\dot{A}}{A}-\frac{\dot{B}}{B}).
\end{equation*}

It is reported in ref. $39$ that, for a spatially homogeneous
universe model, the normal congruence to homogeneous expansion
yields the ratio $\frac{\sigma}{\theta}$ as a constant, i.e., "the
expansion scalar $\theta$ is proportional to the shear scalar
$\sigma$". This physical condition leads to the following relation
between the scale factors:
\begin{equation*}
A=B^{m},
\end{equation*}
where $m\neq1$ is any constant (for $m=1$, it reduces to a flat FRW
model). This condition has been used to investigate various exact
universe models $^{40-42)}$. This assumption leads to the following
set of BD dynamical equations:
\begin{eqnarray}\label{12}
(2m+1)\frac{\dot{B}^{2}}{B^{2}}&=&\frac{\rho}{\phi}+\frac{\omega_{0}}{2}
\frac{\dot{\phi}^{2}}{\phi^{2}}-(m+2)\frac{\dot{B}}{B}\frac{\dot{\phi}}{\phi}
+\frac{V(\phi)}{2\phi},\\\label{13}
2\frac{\ddot{B}}{B}+\frac{\dot{B}^{2}}{B^{2}}&=&-\frac{P}{\phi}-\frac{\omega_{0}}{2}
\frac{\dot{\phi}^{2}}{\phi^{2}}-2\frac{\dot{B}}{B}\frac{\dot{\phi}}{\phi}
-\frac{\ddot{\phi}}{\phi}+\frac{V(\phi)}{2\phi},\\\label{14}
(m+1)\frac{\ddot{B}}{B}+m^{2}\frac{\dot{B}^{2}}{B^{2}}
&=&-\frac{P}{\phi}-\frac{\omega_{0}}{2}
\frac{\dot{\phi}^{2}}{\phi^{2}}-\frac{\ddot{\phi}}{\phi}
-(m+1)\frac{\dot{B}}{B}\frac{\dot{\phi}}{\phi}
+\frac{V(\phi)}{2\phi},\\\label{15}
\ddot{\phi}+(m+2)\frac{\dot{B}}{B}\dot{\phi}&=&\frac{\rho-3P}{2\omega_{0}+3}
-\frac{2}{2\omega_{0}+3}(V-\frac{\phi V_{,\phi}}{2}),\\\label{16}
\dot{\rho}&=&-(m+2)\frac{\dot{B}}{B}(\rho+P).
\end{eqnarray}
Here, the unknown functions are $B,~V,~\rho,~P$ and $\phi$, and
there are only three independent field equations. In order to have a
closed system of equations, we take the power laws for the scale
factor and scalar field as
\begin{equation}\label{17}
B=b_{0}(\frac{t}{t_{0}})^\alpha,\quad\phi=\phi_{0}(\frac{t}{t_{0}})^\beta,
\end{equation}
where $t_{0},~b_{0}$, and $\phi_{0}$ indicate the present time,
scale factor and scalar field, respectively. The deceleration
parameter is given by
\begin{equation*}
q=-(1-\frac{3}{\alpha(m+2)}).
\end{equation*}

For an accelerated expanding solution of the field equations, the
parameter $\alpha$ must be greater than $\frac{m+2}{3}$. Adding
eqs.(\ref{13}) and (\ref{14}), it follows that
\begin{equation}\label{19}
(m+3)\frac{\ddot{B}}{B}+(m^2+1)\frac{\dot{B}^2}{B^2}
=-\omega_{0}(\frac{\dot{\phi}}{\phi})^2-(m+3)\frac{\dot{B}}{B}\frac{\dot{\phi}}{\phi}-2
\frac{\ddot{\phi}}{\phi}-2\frac{P}{\phi}+\frac{V(\phi)}{\phi}.
\end{equation}
Equation (\ref{16}) leads to
\begin{equation}\label{20}
\rho+P=-\frac{\dot{\rho}B}{(m+2)\dot{B}}.
\end{equation}
Subtracting eq.(\ref{12}) from eq.(\ref{19}) and substituting the
power law ansatz given by eq.(\ref{17}) along with eq.(\ref{20}) in
the resulting equation, the energy density for matter can be written
as
\begin{equation}\label{21}
\rho=\rho_{c(m)}t^{\beta-2},
\end{equation}
where
\begin{eqnarray}\nonumber
\rho_{c(m)}&=&\frac{(m+2)\phi_{0}\alpha}{(\beta-2)t_{0}^\beta}[\frac{(m^2
-3m+2)\alpha^2}{2}-\frac{\beta(m\alpha+\alpha+2)}{2}-\frac{(m+3)\alpha}{2}\\\label{22}
&+&(1+\omega_{0})\beta^2].
\end{eqnarray}
Using the energy density given by eq.(\ref{21}) in eq.(\ref{16}),
the corresponding pressure becomes
\begin{equation}\label{23}
P=P_{c(m)}t^{\beta-2},
\end{equation}
where
\begin{eqnarray}\nonumber
P_{c(m)}&=&\frac{\phi_{0}(2-\beta-(m+2)\alpha)}{(\beta-2)t_{0}^\beta}[\frac{(m^2
-3m+2)\alpha^2}{2}-\frac{\beta(m\alpha+\alpha+2)}{2}\\\nonumber
&-&\frac{(m+3)\alpha}{2}+(1+\omega_{0})\beta^2].
\end{eqnarray}
From these expressions, we can write EoS as
\begin{equation}\label{24}
P=\gamma_{B}\rho,\quad\gamma_{B}=\frac{2-\beta}{\alpha(m+2)}-1,
\end{equation}
where $\gamma_{B}$ is the EoS parameter. Since the fluid under
consideration is a perfect fluid $(0<\gamma_{B}<1)$, it constrains
the parameter $\beta$ as
\begin{equation}\label{25}
2-2\alpha(m+2)<\beta<2-\alpha(m+2).
\end{equation}
We would like to mention here that for any positive $m$ other than
$1$ and $\alpha>\frac{3}{m+2}$, $\beta$ remains negative.

Solving the field equations (\ref{12})-(\ref{15}) for the
self-interacting potential, we obtain
\begin{equation}\label{26}
V(\phi)=V_{c}\phi^{\frac{(\beta-2)}{\beta}},
\end{equation}
where
\begin{eqnarray}\nonumber
V_{c}&=&\frac{\phi_{0}^{2/\beta}}{t_{0}^2}[\alpha^2\{\frac{m^2+5m+6}{2}
-(\frac{2\alpha(m+2)-2+\beta}{2-\beta})(\frac{m^2-3m+2}{2})\}\\\nonumber
&+&\alpha(\frac{m+3}{2})\frac{(2\beta+2\alpha(m+2)-4)}{2-\beta}+\beta\{\frac{(3m+7)\alpha}{2}-1+
\frac{(m+1)\alpha+2}{2}\\\label{26*}
&\times&\frac{2\alpha(m+2)-2+\beta}{(2-\beta)}\}
+\beta^2(1-\frac{(1+\omega_{0})(2\alpha(m+2)-2+\beta)}{(2-\beta)})].
\end{eqnarray}
Now, we discuss the energy density and pressure for the missing
energy (dark energy) by taking the energy density and pressure due
to the scalar field. These are given by
\begin{eqnarray}\nonumber
\rho_{\phi}&=&\frac{\omega_{0}}{2}\frac{\dot{\phi}^2}{\phi}+\frac{V}{2}
-\frac{(m+2)\dot{B}\dot{\phi}}{B},\\\nonumber
P_{\phi}&=&\frac{\omega_{0}}{2}\frac{\dot{\phi}^2}{\phi}-\frac{V}{2}
+\ddot{\phi}+\frac{(m+3)\dot{B}\dot{\phi}}{2B}.
\end{eqnarray}
After inserting their respective values, these quantities become
\begin{equation}\label{a}
\rho_{\phi}=\rho_{c(\phi)}t^{\beta-2},\quad
P_{\phi}=P_{c(\phi)}t^{\beta-2},
\end{equation}
where
\begin{eqnarray}\nonumber
\rho_{c(\phi)}&=&[\frac{\omega_{0}\beta^2}{2}-\alpha\beta
(m+2)+\frac{1}{2}
(\alpha^2\{\frac{m^2+5m+6}{2}-(\frac{m^2-3m+2}{2})\\\nonumber
&\times&(\frac{2\alpha(m+2)-2+\beta}{2-\beta})\}+\alpha\frac{(m+3)}{2}\frac{(2\beta
+2\alpha(m+2)-4)}{2-\beta}\\\nonumber
&+&\beta\{\frac{(3m+7)\alpha}{2}-1+\frac{((m+1)\alpha+2)}{2}\frac{(2\alpha(m+2)
-2+\beta)}{(2-\beta)}\}\\\label{28}
&+&\beta^2(1-\frac{(1+\omega_{0})(2\alpha(m+2)-2+\beta)}{(2
-\beta)}))]\frac{\phi_{0}}{t_{0}^\beta},
\end{eqnarray}
\begin{eqnarray}\nonumber
P_{c(\phi)}&=&[\frac{\omega_{0}\beta^2}{2}+\alpha\beta\frac{(m+3)}{2}+\beta^2-\beta-\frac{1}{2}
(\alpha^2\{\frac{m^2+5m+6}{2}-(\frac{m^2-3m}{2}\\\nonumber&+&1)
(\frac{2\alpha(m+2)-2+\beta}{2-\beta})\}
+\alpha\frac{(m+3)}{2}\frac{(2\beta+2\alpha(m+2)-4)}{2-\beta}\\\nonumber
&+&\beta\{\frac{(3m+7)\alpha}{2}-1+\frac{((m+1)\alpha+2)}{2}
\frac{(2\alpha(m+2)-2+\beta)}{(2-\beta)}\}\\\label{29}
&+&\beta^2(1-\frac{(1+\omega_{0})(2\alpha(m+2)-2+\beta)}{(2-\beta)}))]
\frac{\phi_{0}}{t_{0}^\beta}.
\end{eqnarray}
The energy density and pressure for the scalar field obey the EoS,
i.e., $P_\phi=\gamma_{\phi}\rho_{\phi}$, where
\begin{equation}\label{30}
\gamma_{\phi}=-1+\frac{\beta^2(1+\omega_{0})-\beta-\frac{(m+1)}{2}\alpha\beta}{\rho_{c}}.
\end{equation}

In order to have a positive energy density for both matter and
scalar field, the BD parameter should be constrained as
\begin{eqnarray}\nonumber
&&\frac{2(2-\beta)}{\beta^{2}(4-2\beta-2\alpha(m+2))}[\alpha\beta(m+2)
-\frac{(m^2+5m+6)\alpha^2}{4}+(\frac{m^2-3m}{4}\\\nonumber
&&\times\alpha^2+\frac{\alpha^2}{2})\frac{(2\alpha(m+2)
+\beta-2)}{2-\beta}-\frac{\alpha(m+3)(2\alpha(m+2)
+2\beta-4)}{4(2-\beta)}\\\nonumber
&&-\frac{\beta(3m+7)\alpha}{4}+\frac{\beta}{2}-\frac{\beta(2\alpha(m+2)+\beta-2)
(2+\alpha(m+1))}{4(2-\beta)}-\frac{\beta^2}{2}\\\nonumber
&&+\frac{\beta^2(2\alpha(m+2)-2+\beta)}{2(2-\beta)}]<\omega_{0}<[\frac{2+\alpha(m+1)}{2\beta}
+\frac{\alpha(m+3)}{2\beta^2}-1\\\label{31}
&&-\frac{\alpha^2(m^2-3m+2)}{2\beta^2}].
\end{eqnarray}
The deceleration parameter requires $\alpha>\frac{m+2}{3}$ for an
accelerated expansion of the universe. Since $m>0,~(m\neq1)$, we can
discuss two possible values of $m$, i.e., $m>1$ and $0<m<1$. The
feasible region for $\omega_{0}$ allowed by eq.(\ref{31}) as
$(\beta,~\omega_{0})$ space is shown in Fig. \textbf{1}. Fig.
\textbf{1(a)} indicates the $(\beta,~\omega_{0})$ space for $m>1$
and Fig. \textbf{1(b)} shows the allowed region for these parameters
with $0<m<1$. These figures indicate that the positivity condition
for both energy densities implies negative values of the BD
parameter in both cases. We found that the admissible range of BD
parameter turns out to be $-1.5<\omega_{0}<-1$ only for $0<m<1$ with
$\beta<-5.5$. This range for the BD parameter is negative. However,
it does not yield ghost instabilities. A ghost cannot exist
consistently, however, in the present model, the ghost field could
be avoided by a suitable selection of parameters.
\begin{figure}
\centering \epsfig{file=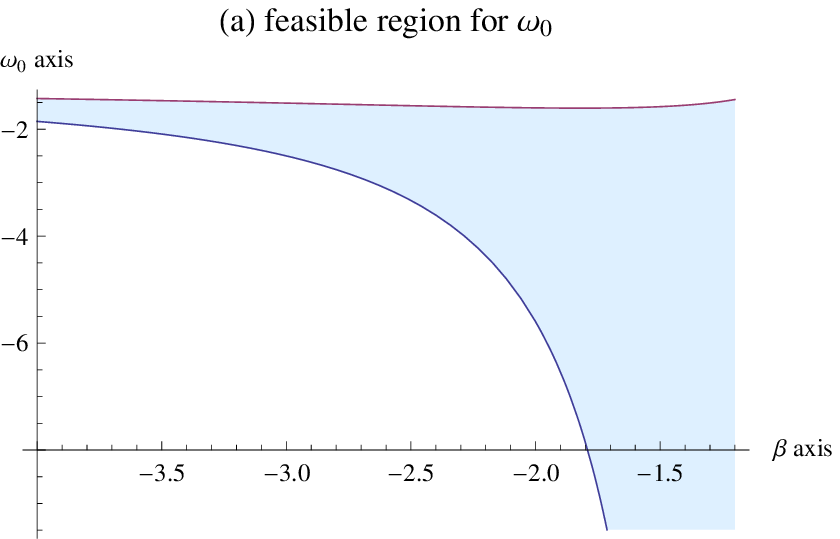,width=.45\linewidth}
\epsfig{file=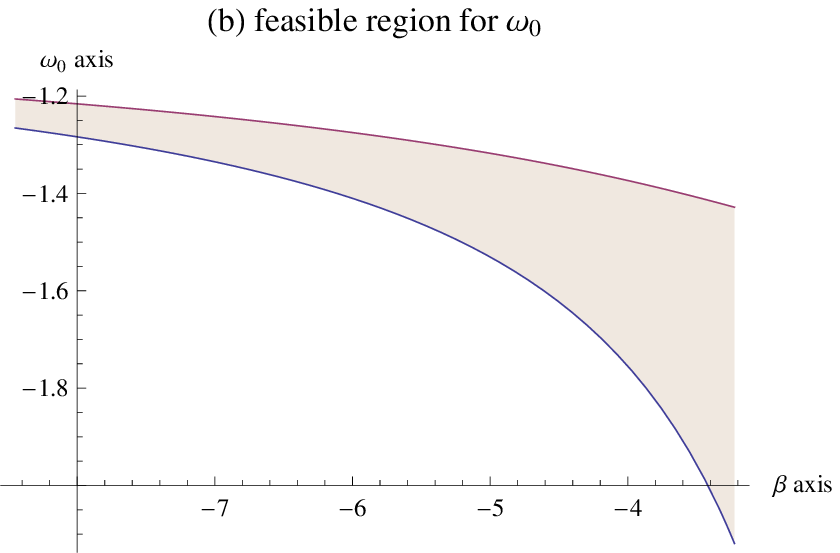,width=.45\linewidth} \caption{Admissible
region for $\omega_{0}$ vs $\beta$ for (a) $m=2$ and $\alpha=0.8$,
(b) $m=3/4$ and $\alpha=1.1$ (Color online).}
\end{figure}

Using eq.(\ref{12}), the matter and scalar field density parameters
are defined as
\begin{equation}\label{32}
\Omega_{m}=\frac{\rho_{m}B^2}{\phi(1+2m)\dot{B}^2},\quad
\Omega_{\phi}=\frac{\rho_{\phi}B^2}{\phi(1+2m)\dot{B}^2}.
\end{equation}
Inserting $B$ and $\phi$ from eq.(\ref{17}), it follows that
\begin{eqnarray}
\Omega_{m}=\frac{\rho_{c(m)}}{(1+2m)\alpha^2},\quad
\Omega_{\phi}=\frac{\rho_{c(\phi)}}{(1+2m)\alpha^2}.
\end{eqnarray}
In terms of matter and the scalar density parameters $\Omega_{m}$
and $\Omega_{\phi}$, the EoS parameter $\gamma_{\phi}$ in
eq.(\ref{30}) can be rewritten as
\begin{equation}\label{33}
\gamma_{\phi}=-1-\frac{\frac{\alpha\beta(1+m)}{2}-\frac{(\beta
-2)t^{\beta}_{0}\alpha(1+2m)\Omega_{m}}{(m+2)\phi_{0}}+\frac{(m^2-3m+2)\alpha^2}{2}
-\frac{(m+3)\alpha}{2}}{(1+2m)\alpha^2\Omega_{\phi}}.
\end{equation}
\begin{figure}
\centering \epsfig{file=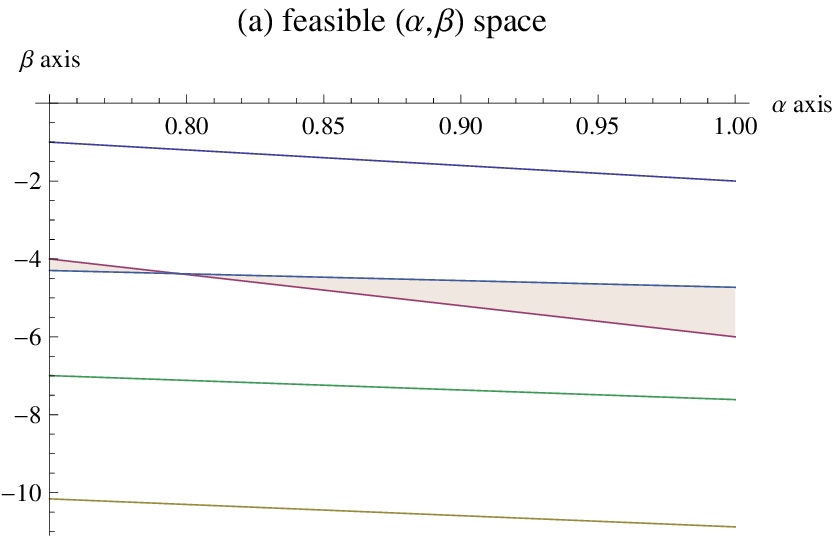,width=.45\linewidth}
\epsfig{file=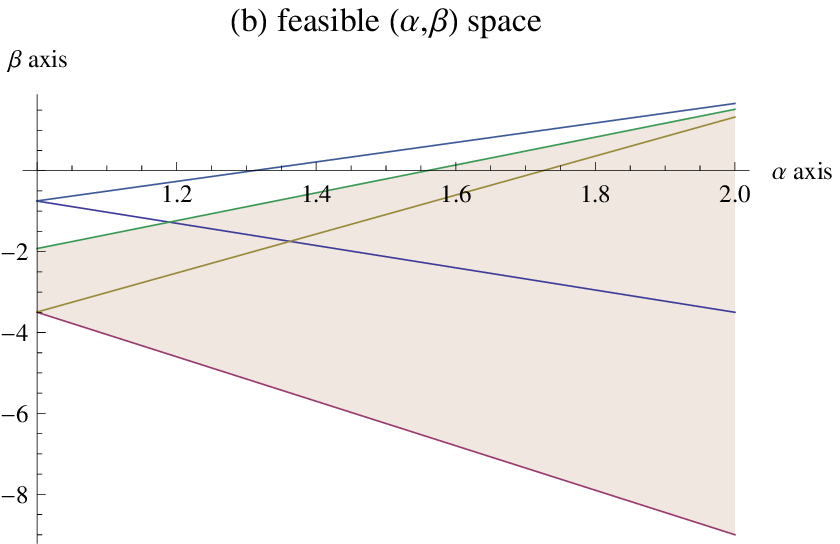,width=.45\linewidth}\caption{Allowed region
for parameter $\beta$ versus $\alpha$ with (a)
$\gamma_{\phi}=-1.24,~m=2$ and (b) $\gamma_{\phi}=-0.6,~m=3/4$
(Color online).}
\end{figure}
\begin{figure}
\centering \epsfig{file=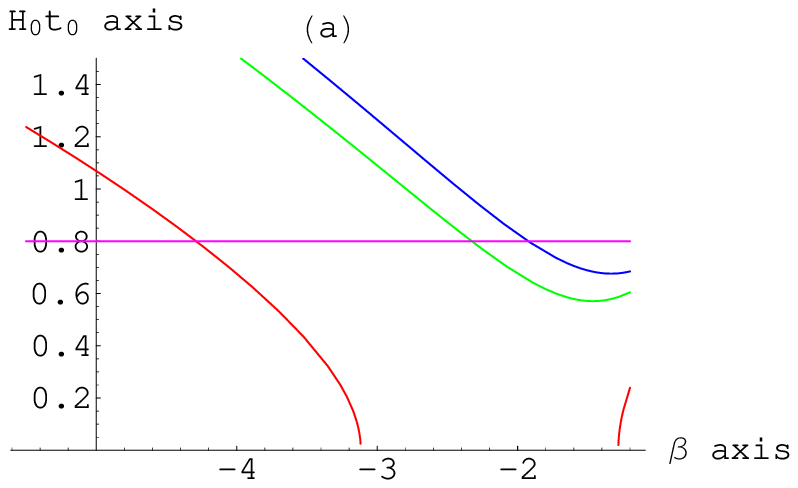,width=.40\linewidth}
\epsfig{file=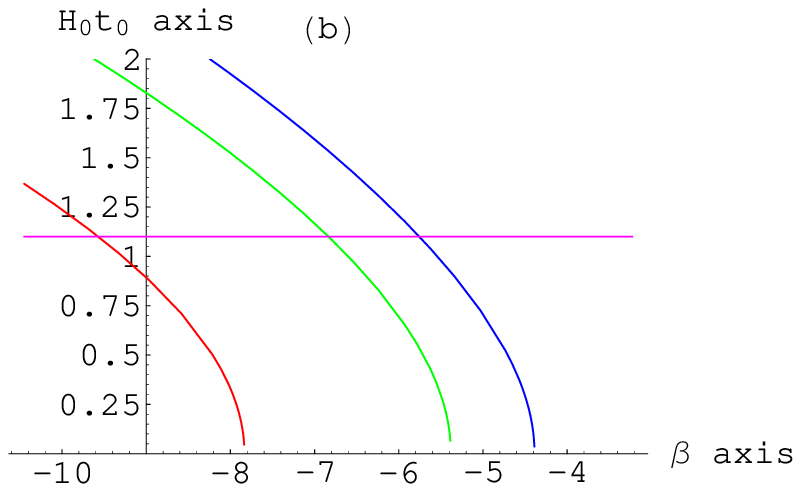,width=.40\linewidth} \caption{(a) Plot of
$H_{0}t_{0}$ vs $\beta$ for $\alpha=0.8,~\Omega_{m}=0.28$, and
$m=2$. Here, the pink horizontal line corresponds to $\alpha=0.8$.
(b) Plot of $H_{0}t_{0}$ vs $\beta$ for
$\alpha=1.1,~\Omega_{m}=0.28$, and $m=3/4$. Here, the pink
horizontal line indicates $\alpha=1.1$ (Color online).}
\end{figure}

The observational data provided by the Supernova Cosmology Project
and High $z$ Survey Project can lead to an accelerated expanding
universe with a positive cosmological constant $^{1,2)}$. By taking
the best-fit values of the density parameters for matter and the
scalar field (for flatness of the universe, i.e.,
$\Omega_{m}=0.28,~\Omega_{\phi}=0.72$) $^{37)}$ as well as the
bounds for $\beta$ given by eq.(\ref{25}), we can determine the
range for the EoS parameter $\gamma_{\phi}$. We consider the case
when $0<m<1$ and take $m=3/4$. The corresponding constraints for the
other parameters are $\alpha>1.09$ and $-8.45<\beta<-3.225$. Using
these values, the constraint for the EoS parameter $\gamma_{\phi}$
turns out to be $-1.44<\gamma_{\phi}<0.71$. Likewise, in the case of
$m>1$, we take $m=2$ and thus we have $\alpha>0.75$ and
$-4.4<\beta<-1.2$. The corresponding bounds for $\gamma_{\phi}$ will
be $-1.8<\gamma_{\phi}<-0.23$. The allowed region in the
$(\alpha,~\beta)$ space with $0.2<\Omega_{m}<0.4$, which satisfies
$-1.44<\gamma_{\phi}<0.71$ and $-1.8<\gamma_{\phi}<-0.23$, is shown
as a shaded region in Fig. \textbf{2}. In Fig. \textbf{2(a)}, we
take the case $m>1$, and Fig. \textbf{2(b)} corresponds to the case
$0<m<1$.

As the BD theory of gravity has a salient feature of a time-varying
gravitational constant, i.e., $G(t)=\frac{1}{\phi(t)}$, the rate of
change in the gravitational constant at the present time can be
calculated as
\begin{equation}\label{34}
[\frac{\dot{G}}{G}]_{t=t_0}=-\frac{\beta}{\alpha}H_{0},
\end{equation}
where $H_{0}$ is the directional Hubble parameter
$\frac{\dot{B}}{B}$ at the present time and we have used the power
law ansatz eq.(\ref{17}). For a viable model representing the
accelerated expansion of the universe, the constraint
$[\frac{\dot{G}}{G}]_{t=t_0}<10^{-10}$ should be satisfied $^{39)}$.
For our model, the best-fit value of the parameter
$\alpha=H_{0}t_{0}$ admitting the accelerated expansion can be
written as
\begin{eqnarray}\nonumber
(H_{0}t_{0})^2&=&\frac{\alpha(m+2)}{(1+2m)\Omega_{m}}[\frac{m^2-3m+2}{2}\alpha^2
-\frac{\beta(m\alpha+\alpha+2)}{2}-\frac{(m+3)\alpha}{2}\\\label{35}
&+&(1+\omega_{0})\beta^2].
\end{eqnarray}
The graphical behavior of this expression in the cases $0<m<1$ and
$m>1$ along with appropriate values of other parameters is shown in
Fig. \textbf{3}. It can be seen that the graphs of $H_{0}t_{0}$ with
the allowed range of $\omega_{0}$ intersect at least at one point on
the respective horizontal lines $\alpha=0.8$ and $\alpha=1.1$ in
both cases.

The correction to Newton's law leads to two kind of problems: the
variation in the gravitational constant and the propagation of the
fifth force. From the cosmological point of view, the gravitational
constant should be positive (it restricts $\phi_{0}$ to be positive)
and its variation should be constrained within certain limits
$^{13-15)}$. The case of a constant BD parameter yields small
variations that should be constrained by
$[\frac{\dot{G}}{G}]_{0}<4\times10^{-10}$ yrs. In our case, the
expression is given by $[\frac{\dot{G}}{G}]_0=-\frac{\beta}{t_0}$,
where the present time is taken to be $t_{0}=14\pm2$ Gyr. By
choosing the appropriate values of the parameters, it can be shown
that in our case this constraint can be easily satisfied. For
example, if we take $\alpha=0.8,~m=2$ and $\beta=-1.3$ as
$-4.4<\beta<-1.2$, then we have
$[\frac{\dot{G}}{G}]_{0}=0.93\times10^{-10}$, which safely lies
within the limit $<4\times10^{-10}$. Thus it can be concluded that,
in our case, the variation in the gravitational constant (due to
presence of scalar field) could satisfy the observational limit
suggested for cosmic acceleration.

The Brans-Dicke theory in the Jordan frame can be transformed to the
Einstein frame by defining a conformal transformation. The resulting
action in the Einstein frame corresponds to a coupled quintessence
field scenario with the coupling given by
$2Q^2=\frac{1}{2\omega_{BD}+3}$ (this leads to a constant coupling
$Q=-\frac{1}{\sqrt{6}}$ in metric $f(R)$ gravity) $^{43-47)}$. It is
argued that when this coupling is of order unity, local gravity
tests become inconsistent owing to the propagation of fifth force
between the field and the non relativistic matter. Whenever the
gravity is modified from GR, there are some constraints that come
from the local gravity tests like solar system tests, and the
violation of the equivalence principle etc.

The action of the BD theory in the Jordan frame can be transformed
to the Einstein frame via conformal transformation defined by the
coupling factor $F=e^{-2Q\phi}$ (basically, we define
$g_{\mu\nu}=F(\phi)g_{\mu\nu}$). For a massless scalar field, the
coupling factor should be constrained by the condition
$|Q|<2.5\times10^{-3}$ that comes from the experimental bound on the
BD parameter $\omega_{BD}>40,000$. When the field is massive, this
coupling factor can be large $|Q|\sim1$, consequently, the Chameleon
mechanism provides a way to be consistent with local gravity tests.
In this regard, the field mass needs to be small in order to realize
the present acceleration on cosmological scales $m_{\phi}\approx
H_{0}$, while it is large in the region of high density to avoid the
propagation of the fifth force $m_{\phi}\gg H_{0}$. The field that
changes its mass depending on the environment is the Chameleon
field. In a spherically symmetric configuration, the field equation
in the Einstein frame is given by
\begin{eqnarray}\nonumber
\frac{d^2\phi}{dr^2}+\frac{2}{r}\frac{d\phi}{dr}-\frac{dU_{eff}}{d\phi}=0,
\end{eqnarray}
where $U_{eff}=U+e^{Q\phi}\rho_{m}$. A detailed discussion on thick-
and thin-shell solutions in this context by finding the minima of
the effective potential is available in the literature $^{43-47)}$.

The thin-shell parameter is defined to be
$\epsilon_{th}=\frac{\phi_{B}-\phi_{A}}{6Q\phi_{c}}$ and its upper
bound can be found using solar system tests or the violation of the
equivalence principle. If the thin-shell parameter is much less than
1, i.e., $\epsilon_{th}<\frac{1.9\times10^{-5}}{Q^2}$, then solar
system tests are satisfied even for $|Q|=O(1)$. Likewise, the
propagation of the fifth force (defined by
$F=-6Q^2\epsilon_{th}\frac{GM_{c}}{r}$) will be suppressed as
compared to gravitational force if
$\epsilon_{th}<\frac{8.8\times10^{-7}}{Q^2}$, which further leads to
the constraint $|\phi_{B}|<3.7\times10^{-15}$, where
$|\phi_{B}|\gg|\phi_{A}|$. This analysis is then applied to a
runaway form of the potential, e.g., the inverse power law
potential, and the respective constraint on $M$ is discussed
$^{45,46)}$. In our case, we have found the self interacting
potential with a positive power as
$V=V_{c}\phi^{\frac{\beta-2}{\beta}},~\frac{\beta-2}{\beta}>0$, for
negative $\beta$ values. In the present case, the minima of the
effective potential inside and outside a spherically symmetric body
($\rho_{A}\gg\rho_{B}$) are given by
\begin{eqnarray}\nonumber
\phi_{A}=(\frac{-nV_{c}}{Q\rho_{A}})^{1/1-n}, \quad
\phi_{B}=(\frac{-nV_{c}}{Q\rho_{B}})^{1/1-n}.
\end{eqnarray}
In order to suppress the propagation of the fifth force, we must
have $-nV_{c}<(3.7\times10^{-15})^{1-n}Q\rho_{B}$, where $\rho_{B}$
can be taken as $10^{-24}g/cm^3$. Moreover, for the Chameleon
mechanism to work, the mass of the field should be constrained by
the condition $m_{A}>H_{0}$ (this allows the condition
$1/(m_{A}r_{c})\ll1$) which in our case results
$\frac{(n-1)(-Q\rho_{A})^{\frac{2-n}{1-n}}}{(nV_{c})^{1/1-n}}<H_{0}$.

\section{Energy Density Perturbations}

In this section, we investigate whether the issue of structure
formation is modified by the dynamics of the BD scalar field within
the BI universe model. This can be achieved by considering the
evolution of energy density perturbations. Basically, we perturb the
field equations and take the temporal components of these equations.
The behavior of the relevant variables for the accelerated expansion
of the universe is investigated. For this purpose, we introduce the
notations $\delta g_{\mu\nu}=h_{\mu\nu},~\delta R_{00},~\delta
u^\mu$, and $\delta T^{00}$ to denote the perturbations in
$g_{\mu\nu},~R_{00},~u^\mu$, and $T^{00}$, respectively. For such
perturbations, we use a synchronous coordinate condition given by
$h_{\mu 0}=0,~\forall\mu$. Using perturbation in the metric tensor
defined by
\begin{equation*}
\overline{g}_{\mu\nu}=g_{\mu\nu}+h_{\mu\nu},
\end{equation*}
the perturbed form of the Ricci tensor can be written as
\begin{equation}\label{40}
\delta
R_{\mu\nu}=\chi^{\rho}_{\mu\nu;\rho}-\chi^{\rho}_{\mu\rho;\nu},
\end{equation}
where
\begin{equation*}
\chi^{\rho}_{\mu\nu}=\frac{g^{\rho\sigma}}{2}[h_{\sigma\mu;\nu}
+h_{\sigma\nu;\mu}-h_{\mu\nu;\sigma}].
\end{equation*}
The non-zero temporal component of the Ricci tensor is
\begin{equation}\label{41}
R_{00}=-\frac{(m+2)\ddot{B}}{B}+m(1-m)\frac{\dot{B}^2}{B^2}.
\end{equation}
Multiplying eq.(\ref{19}) by $\frac{(m+2)}{(m+3)}$ and after some
manipulation, it follows that
\begin{eqnarray}\nonumber
R_{00}&=&\omega_{0}\frac{\dot{\phi}^2}{\phi^2}+\frac{\dot{\phi}}{\phi}
+\frac{\rho}{\phi}[\frac{(m+1)(1-3\gamma_{B})+2(m+2)(2\omega_{0}
+3)\gamma_{B}}{(m+3)(2\omega_{0}+3)}\\\label{42}
&+&\frac{2}{(m+3)}]\frac{2(1+m)}{(2\omega_{0}+3)(m+3)}[\frac{V}{\phi}
-\frac{V_{,\phi}}{2\phi}]-\frac{(m+1)V}{(m+3)\phi}.
\end{eqnarray}
Using the definition of the perturbed Ricci tensor given in
eq.(\ref{40}), the perturbed temporal component of the Ricci tensor
is determined by
\begin{eqnarray}\nonumber
\delta
R_{00}&=&\frac{1}{2B^{2m}}[\ddot{h}_{11}-4m\frac{\dot{B}}{B}\dot{h}_{11}
+2m((1+2m)\frac{\dot{B}^2}{B^2}-\frac{\ddot{B}}{B})h_{11}]\\\label{43}
&+&\frac{1}{2B^2}[\ddot{h}_{kk}-4\frac{\dot{B}}{B}\dot{h}_{kk}
+2(3\frac{\dot{B}^2}{B^2}-\frac{\ddot{B}}{B})h_{kk}],\quad
k=0,2,3.
\end{eqnarray}
The trace of the energy-momentum tensor is perturbed as
\begin{equation}\label{44}
\overline{T}=T+\delta T=T+\delta\rho-3\delta P.
\end{equation}
The action of the perturbed D' Alembertain operator on the BD field
is
\begin{eqnarray}\nonumber
\delta(\Box\phi)&=&\delta\ddot{\phi}+(m+2)\frac{\dot{B}}{B}\delta\dot{\phi}
+\dot{\phi}[-\frac{1}{2}(\frac{\dot{h}_{11}}{B^{2m}})
-\frac{1}{2}(\frac{\dot{h}_{kk}}{B^{2}}]\\\label{44}
&-&\frac{1}{B^{2m}}\frac{\partial^2}{\partial^2
x}(\delta\phi)-\frac{1}{B^2}\frac{\partial^2}{\partial^2 y}
(\delta\phi)-\frac{1}{B^2}\frac{\partial^2}{\partial^2
z}(\delta\phi).
\end{eqnarray}
We define he following parameters for relevant perturbations
\begin{eqnarray}\label{45}
h_{11}=B^{2m}h,\quad h_{kk}=B^2h,\quad
\delta\phi=\lambda\phi,\quad \lambda\ll
1,\quad\delta\rho=\Delta\rho,\quad \Delta\ll 1.
\end{eqnarray}
Here, $h(t),~\lambda(t)$, and $\Delta(t)$ denote the perturbed
gravitational, scalar and matter density fields, respectively. Using
these parameters in independent field equations, we obtain
\begin{eqnarray}\nonumber
&&\ddot{h}=\ddot{\lambda}+2(1+\omega_{0})\dot{\lambda}\frac{\dot{\phi}}{\phi}
+(\Delta-\lambda)\frac{\rho}{\phi}[\frac{(m+1)(1-3\gamma_{B})}{(m+3)(2\omega_{0}+3)}\\\nonumber
&&+\frac{2(m+2)(2\omega_{0}+3)\gamma_{B}}{(m+3)(2\omega_{0}+3)}+\frac{2}{(m+3)}]
-\frac{(m+1)}{(m+3)}[1-\frac{(2-n_{1})}{(2\omega_{0}+3)}]V_{c}(n_{1}\\\label{48}
&&-1)\lambda\phi^{n_{1}-1},\\\nonumber
&&\ddot{\lambda}+\dot{\lambda}(2\frac{\dot{\phi}}{\phi}+(m+2)\frac{\dot{B}}{B})
+\lambda(\frac{\ddot{\phi}}{\phi}+(m+2)\frac{\dot{B}}{B}\frac{\dot{\phi}}{\phi})
-\frac{\dot{h}\dot{\phi}}{\phi}-\frac{1}{B^{2m}}\frac{\partial^2\lambda}{\partial^2x}\\\label{49}
&&-\frac{1}{B^2}(\frac{\partial^2\lambda}{\partial^2y}+
\frac{\partial^2\lambda}{\partial^2z})=\frac{\Delta\rho(1
-3\gamma_{B})}{\phi(2\omega_{0}+3)}-\frac{(2-n_{1})V_{c}}{(2\omega_{0}+3)},\\\label{50}
&&\dot{\Delta}-\dot{h}+\delta u^{i}_{,i}=0.
\end{eqnarray}
Here, $u^{i}$ is the comoving velocity of the fluid and $\delta
u^{i}_{,i}$ denotes the respective perturbed quantity. We also
introduce the notation $n_{1}=\frac{\beta-2}{\beta}$. For the
discussion of structure formation (as the structure is formed in the
absence of pressure, which can prevent gravitational collapse), we
consider a pressureless case, i.e., $\gamma_{B}=0$. Thus, the
parameters $\alpha$ and $\beta$ turn out to be
\begin{equation}\label{51}
\beta=\frac{2}{1-n_{1}},\quad
\alpha=\frac{-2n_{1}}{(1-n_{1})(m+2)}.
\end{equation}

By using an infinitesimal gauge transformation, we can take the
perturbed four-velocity as null, for which eq.(\ref{50}) leads to
\begin{equation}\label{52}
\Delta=h,
\end{equation}
where the constant of integration is taken to be zero. We also
consider a plane wave-like behavior of the considered perturbations,
that is,
\begin{eqnarray}\label{53}
\lambda(\textbf{x},t)=\lambda(t)\exp(-\iota
\textbf{k}.\textbf{x}),
\end{eqnarray}
where $\textbf{k}$ is the wave number. Substituting eqs.(\ref{17})
and (\ref{53}) into eqs.(\ref{48}) and (\ref{49}), it follows that
\begin{eqnarray}\nonumber
\ddot{\Delta}&=&\ddot{\lambda}+\frac{4(1+\omega_{0})\dot{\lambda}}{(1-n_{1})t}
+\frac{(\Delta-\lambda)}{t^2}\overline{\rho_{c}}[\frac{(m+1)
+2(2\omega_{0}+3)}{(m+3)(2\omega_{0}+3)}]\\\label{54}
&-&\frac{(m+1)}{(m+3)}V_{c}[1-\frac{2-n_{1}}
{2\omega_{0}+3}](n_{1}-1)\lambda\frac{\phi_{0}^{(n_{1}
-1)}t^2_{0}}{t^2},
\end{eqnarray}
\begin{eqnarray}\nonumber \ddot{\lambda}&=&-\frac{\dot{\lambda}}{t}
\frac{(4-2n_{1})}{(1-n_{1})}-\frac{2\lambda}{t^2(1-n_{1})}
+\frac{2\dot{\Delta}}{t(1-n_{1})}+\frac{\lambda
k_{1}^2}{B_{0}^{2m}}(\frac{t}{t_{0}})^{\frac{4mn_{1}}{(1-n_{1})(m+2)}}\\\nonumber
&+&\frac{\lambda(k_{2}^2+k_{3}^2)}{B_{0}^2}(\frac{t}{t_{0}})^{\frac{4n_{1}}{(1
-n_{1})(m+2)}}+\frac{\Delta\overline{\rho_{c}}}{(2\omega_{0}+3)t^2}-\frac{(2-n_{1})t_{0}^2}
{(2\omega_{0}+3)t^2}V_{c}\lambda
n_{1}\phi_{0}^{(n_{1}-1)},\\\label{55}
\end{eqnarray}
where
$\overline{\rho}_{c}=\frac{\rho_{c(m)}}{\phi_{0}}t_{0}^{2/(1-n_{1})}$.
The expressions for $\rho_{c(m)}$ and $V_{c}$ in terms of $n_{1}$
are
\begin{eqnarray}\nonumber
\rho_{c(m)}&=&-\frac{\phi_{0}}{t_{0}^{2/(1-n_{1})}}
[\frac{m^2-3m+2}{2}\frac{4n_{1}^2}{(1-n_{1})^2(m+2)^2}
+\frac{2n_{1}(m+1)}{(1-n_{1})^2(m+2)}\\\label{56}
&-&\frac{2}{1-n_{1}}+\frac{(m
+3)n_{1}}{(1-n_{1})(m+2)}+\frac{4(1+\omega_{0})}{(1-n_{1})^2}],\\\label{57}
V_{c}&=&\frac{\phi_{0}^{(1-n_{1})}}{t_{0}^2}[\frac{8n_{1}^2(1+2m)}{(m+2)^2(1
-n_{1})^2}-\frac{4n_{1}}{(1-n_{1})^2}-\frac{4\omega_{0}}{(1-n_{1})^2}].
\end{eqnarray}

Combining eqs.(\ref{54}) and (\ref{55}) and neglecting the terms
involving higher powers of $t^{-2}$, we have
\begin{equation}\label{58}
\ddot{\Delta}+\frac{C_{1}}{t}(\dot{\Delta}-\dot{\lambda})+\frac{C_{2}}{t^2}(\Delta
-\lambda)+\frac{C_{3}\dot{\lambda}}{t}+\frac{c_{4}\lambda}{t^2}=0,
\end{equation}
where
\begin{eqnarray}\nonumber
C_{1}(\omega_{0})&=&-\frac{2}{1-n_{1}},\\\nonumber
C_{2}(\omega_{0})&=&[\frac{2(m+2)+2(2\omega_{0}+3)}{(m+3)(2\omega_{0}+3)}][\frac{m^2-3m+2}{2}
\frac{4n_{1}^2}{(1-n_{1})^2(m+2)^2}\\\nonumber
&+&\frac{2n_{1}(m+1)}{(1-n_{1})^2(m+2)}-\frac{2}{1-n_{1}}+\frac{(m
+3)n_{1}}{(1-n_{1})(m+2)}+\frac{4(1+\omega_{0})}{(1-n_{1})^2}],\\\nonumber
C_{3}(\omega_{0})&=&C_{1}(\omega_{0})+\frac{(4-2n_{1})}{(1-n_{1})}
-\frac{4(1+\omega_{0})}{(1-n_{1})},\\\nonumber
C_{4}(\omega_{0})&=&C_{2}(\omega_{0})+\frac{2}{1-n_{1}}
+\frac{(m+1+2(2\omega_{0}+3))}{(m+3)(2\omega_{0}+3)}\overline{\rho}_{c}
+[\frac{(2-n_{1})n_{1}}{2\omega_{0}+3}\\\nonumber
&+&\frac{m+1}{m+3}(1-n_{1})(1-\frac{2
-n_{1}}{2\omega_{0}+3})]V_{c}\phi_{0}^{(n_{1}-1)}t_{0}^2.
\end{eqnarray}
For the solution of eq.(\ref{58}), we take the following
assumptions:
\begin{equation}
\Delta-\lambda=f(t),\quad f(t)=\xi t^\delta,\quad \Delta=\chi
t^\theta,
\end{equation}
where $\xi$ and $\chi$ are arbitrary constants. The substitution of
the above assumption in eq.(\ref{58}) leads to $\delta=\theta$ and a
quadratic equation for $\theta$ given by
\begin{eqnarray}\label{59}
\chi
\theta^2+\theta[\chi(C_{3}-1)+\xi(C_{1}-C_{3})]+C_{4}(\chi-\xi)+C_{2}\xi=0.
\end{eqnarray}
The roots of this equation are
\begin{eqnarray}\nonumber
\theta_{\pm}&=&\frac{1}{2\chi}[(\chi(1-C_{3})+\xi(C_{3}-C_{1}))\\\label{60}
&\pm&\sqrt{(\chi(1-C_{3})+\xi(C_{3}-C_{1}))^2
-4\chi((C_{2}-C_{4})\xi-C_{4}\xi)}].
\end{eqnarray}

In previous papers $^{13,37)}$, the growing modes for density
perturbations are calculated in the asymptotic limit of
$|\omega_{0}|\gg1$. In a similar way, the solution $\theta$ in our
case turns out to be
\begin{eqnarray}\nonumber
\theta_{\pm}&\rightarrow&
\frac{2\omega_{0}}{(1-n_{1})}[(1-\frac{\xi}{\chi})\\\label{61}
&\pm&\sqrt{(1-\frac{\xi}{\chi})^2
+\frac{1}{(m+3)}[(m+1)(1-n_{1})(1-\frac{\xi}{\chi})-2\frac{\xi}{\chi}]}].
\end{eqnarray}
This shows that in the limit $|\omega_{0}|\gg1$, the asymptotic
value of $\theta$ depends on the power of the self-interacting
potential $n_{1}$ and the parameter $m$ for scale factors. As
$\beta$ always remains negative, $n_{1}=\frac{\beta-2}{\beta}$ will
always remain positive and greater than 1. Also, $m$ is a positive
constant with $m\neq1$ implying that $\theta_{+}$ represents the
growing mode for $\xi<\chi$ in both cases $0<m<1$ and $m>1$. Note
that for $m=1$, the modes we obtained match with those obtained by
Sen and Seshadri $^{37)}$ and $n_{1}=2$ corresponds to the modes
obtained by Bertolami and Martins $^{13)}$. Thus, density
perturbations can grow with time using the admissible negative range
of $\omega_{0}$ (found in the previous section) and appropriate
choices for the constants $\xi,~\chi,$ and $n_{1}$ as well as for
$m$ in the asymptotic limits. It can be confirmed that the growing
modes can also be allowed for positive BD parameter values if we
take $\beta>0$ or set the arbitrary constants $\xi>\chi$. However,
the first possibility remains incompatible with the accelerated
expansion of the model (as it can be obtained by setting
$\alpha<3/(m+2)$ with $q>0$).

\section{Summary and Discussion}

In this study, we have investigated the nature of the
self-interacting potential and dark energy, i.e., the energy density
of the BD scalar field for the BI universe model. For this purpose,
we have considered the power law ansatz for the scale factors as a
well as the scalar field. By taking the matter contents as perfect
fluid, we find constraints for all the relevant parameters and their
best-fit values. The feasible regions for these parameters are shown
graphically for the two cases $m>1$ and $0<m<1$. We have also
discussed the variation in the gravitational constant and the
evolution of energy density perturbations for the obtained model.
The results are summarized as follows.
\begin{itemize}
\item The obtained model is entirely based on the parameters
$\alpha,~\beta,~\omega_{0}$, and $m$. Using the deceleration
parameter, $\alpha>\frac{(m+2)}{3}$ for an accelerated expanding
universe model. On the basis of $\alpha$, the bounds for $\beta$ are
found which indicate that $\beta$ always remains negative for
$\alpha>\frac{m+2}{3}$ and $m\neq1,~m>0$. By imposing positivity
condition of energy density for both matter and scalar field, we
find bounds for the BD parameter depending on these parameters. It
is seen that for admissible ranges of these parameters, we have
$-2\leq\omega_{0}\leq-1.4$ and $-1.45\leq\omega_{0}\leq-1.22$ for
$m>1$ and $0<m<1$, respectively.
\item The bounds for $\beta$ provide a positive power law potential,
which can be used to derive the accelerated cosmic expansion.
\item Using the best-fit values for the flatness of the universe
$\Omega_{m}$ and $\Omega_{\phi}$, we find the ranges for the EoS
parameter $\gamma_{\phi}$. For $0<m<1$, this parameter is
constrained as $-8.45<\beta<-3.225$, and for $m>1$, as
$-1.8<\gamma_{\phi}<-0.23$. These ranges are compatible with the
ranges suggested by the observational data $^{48)}$.
\item With the help of density perturbation analysis, it is shown
that $\theta_{+}$ represents the growing mode for suitable choices
of parameters in both cases, $0<m<1$ and $m>1$. Thus, it can be
concluded that the modifications due to the dynamics of BD field do
not disturb structure formation for the BI universe model.
\end{itemize}

Note that the BD constant coupling parameter has a negative range
obtained by imposing the positive energy condition and is obviously
inconsistent with solar system experiment bounds. It can be
confirmed that, if we take negative range of $m$, e.g., $-2<m<0$,
then we can have positive values of the BD parameter, which leads to
the failure of the positive energy condition $^{13)}$. According to
recent observations, the failure of the positive energy condition is
found to be more serious as the contribution of DE must be dominant
as compared to matter density. This problem has widely been observed
in the context of scalar tensor theories, e.g., $^{14,37,49,50)}$.
Recently, it has been found $^{51)}$ that the accelerated expansion
is incompatible with large values of the BD parameter by showing
that scalar-vacuum BD equations in 5D are equivalent to the BD
theory in 4D with a self-interacting potential and an effective
matter field. Moreover, it has been pointed out $^{52)}$ that
structure formation also remains incompatible with solar system
experiment bounds.

The solar system experiments constrain the BD parameter to be
greater than 40,000 for a massless scalar field, but it is not true
when the scalar field is massive. In this case, we define $m\gg
r^{-1}$, where $r$ is the experiment scale and $m$ is the mass of
potential as the spatial dynamics of scalar field can freeze for
this limit. The mass scale can be taken as $m_{AU}=10^{-27}$, where
the relevant scale is taken to be the Astronomical unit
$(1AU=10^8km)$. By comparing with this mass scale, we say that the
scalar field is massive $(m\gg m_{AU})$ or has negligible mass
$(m\ll m_{AU})$. In our case, the self-interacting potential is
$V=V_{c}\phi^{\frac{\beta-2}{\beta}}$, with $V_{c}$ given by
eq.(\ref{26*}), $t_{0}$ is the age of the universe and
$\frac{\beta-2}{\beta}>0$. If the mass of the potential term, i.e.,
$V_{,\phi\phi}$, is greater than the mass scale, then the scalar
field is massive, consequently, it could be expected that all the
values of the BD parameter are admissible as shown in ref. $53$,
where the potential term is taken to be $V=\lambda\phi^{n};~n>0$,
which is quite similar to our case when $n=\frac{\beta-2}{\beta}$
and $\lambda=V_{c}$. However, an exact estimation could be found
using the solar system constraints of the Post-Newtonian parameter
as shown in the literature $^{54)}$. In that case, the potential is
taken to be $U(\phi)=\frac{1}{2}m^2(\phi-\phi_{0})^2$, which is a
subcase of the potential derived in the present study for $\beta=-2$
and it is shown that, for $m>200m_{AU}$, all values of $\omega>-3/2$
are admissible at a $2\sigma$ confidence level.

In our recent work $^{31)}$, we have studied the ``early- and
late-time" behaviors of a BI universe model using perfect,
anisotropic and magnetized anisotropic matter contents. In the
current study, we impose the condition $\alpha>\frac{3}{m+2}$ (which
is necessary for a negative deceleration parameter) and to determine
will be the nature of the self-interacting potential $V$, and the
energy densities and pressures due to matter and the scalar field.
By exploring the admissible range of the BD coupling parameter and
consequently, the EoS parameters $\gamma_{B}$ and $\gamma_{\phi}$,
it is concluded that these values are well-consistent with the
observed values.

\vspace{0.5cm}

{\bf Acknowledgement}

\vspace{0.5cm}

The authors are grateful to the Physical Society Japan for the
financial support in publication.

\vspace{0.5cm}

1) S. Perlmutter, S. Gabi, G. Goldhaber, A. Goobar, D. E. Groom, I.
M. Hook, A. G. Kim, M. Y. Kim, G. C. Lee, R. Pain, C. R.
Pennypacker, I. A. Small, R. S. Ellis, R. G. McMahon, B. J. Boyle,
P. S. Bunclark, D. Carter, M. J. Irwin, K. Glazebrook, H. J. M.
Newberg, A. V. Filippenko, T. Matheson, M. Dopita and W. C. Couch:
Astrophys. J. \textbf{483} (1997) 565; S. Perlmutter, G. Aldering,
M. D. Valle, S. Deustua, R. S. Ellis, S. Fabbro, A. Fruchter, G.
Goldhaber, A. Goobar, D. E. Groom, I. M. Hook, A. G. Kim, M. Y. Kim,
R. A. Knop, C. Lidman, R. G. McMahon, P. Nugent, R. Pain, N.
Panagia, C. R. Pennypacker, P. Ruiz-Lapuente, B. Schaefer and N.
Walton: Nature \textbf{391} (1998) 51; S. Perlmutter, G. Aldering,
G. Goldhaber, R. A. Knop, P. Nugent, P. G. Castro, S. Deustua, S.
Fabbro, A. Goobar, D. E. Groom, I. M. Hook, A. G. Kim, M. Y. Kim, J.
C. Lee, N. J. Nunes, R. Pain, C. R. Pennypacker, R. Quimbey, C.
Lidman, R. S. Ellis, M. Irwin, R. G. Mcmahon, P. Ruiz-lapuente, N.
Walton, B. Schaefer, B. J. Boyle, A. V. Filippenko, T. Matheson, A.
S. Fruchter, N. Panagia, H. J. M. Newberg, and
W. J. Couch: Astrophys. J. \textbf{517} (1999) 565.\\
2) A. G. Riess, A. V. Filippenko, P. Challis, A. ClocChiatti, A.
Diercks, P. M. Garnavich, R. L. Gilliland, C. J. Hogan, S. Jha, R.
P. Kirshner, B. Leibundgut, M. M. Phillips, D. Reiss, B. P.
Schmidt, R. A. Schommer, R. C. Smith, J. Spyromilio, C. Stubbs, N.
B. Suntzeff, and J. Tonry: Astron. J. \textbf{116} (1998) 1009.\\
3) C. L. Bennett, M. Halpern, G. Hinshaw, N. Jarosik, A. Kogut, M.
Limon, S. S. Meyer, L. Page, D. N. Spergel, G. S. Tucker, E.
Wollack, E. L. Wright, C. Barnes, M. R. Greason, R. S. Hill, E.
Komatsu, M. R. Nolta, N. Odegard, H. V. Peiris, L. Verde, and
J. L. Weiland: Astrophys. J. Suppl. \textbf{148} (2003) 1.\\
4) M. Tegmark, M. A. Strauss, M. R. Blanton, K. Abazajian, S.
Dodelson, H. Sandvik, X. Wang, D. H. Weinberg, I. Zehavi, N. A.
Bahcall, F. Hoyle, D. Schlegel, R. Scoccimarro, M. S. Vogeley, A.
Berlind, T. Budavari, A. Connolly, D. J. Eisenstein, D.
Finkbeiner, J. A. Frieman, J. E. Gunn, L. Hui, B. Jain, D.
Johnston, S. Kent, H. Lin, R. Nakajima, R. C. Nichol, J. P.
Ostriker, A. Pope, R. Scranton, U. Seljak, R. K. Sheth, A.
Stebbins, A. S. Szalay, I. Szapudi, Y. Xu, J. Annis, J. Brinkmann,
S. Burles, F. J. Castander, I. Csabai, J. Loveday, M. Doi, M.
Fukugita, B. Gillespie, G. Hennessy, D. W. Hogg, Z. E. Ivezic´, G.
R. Knapp, D. Q. Lamb, B. C. Lee, R. H. Lupton, T. A. McKay, P.
Kunszt, J. A. Munn, L. Connell, J. Peoples, J. R. Pier, M.
Richmond, C. Rockosi, D. P. Schneider, C. Stoughton, D. L. Tucker,
D. E. V. Berk, B. Yanny, and D. G. York: Phys. Rev. D \textbf{69} (2004) 03501.\\
5) S. W. Allen, R. W. Schmidt, H. Ebeling, A. C. Fabian, and L. V.
Speybroeck: Mon. Not. Roy. Astron. Soc.
\textbf{353} (2004) 457.\\
6) E. Hawkins, S. Maddox, S. Cole, O. Lahav, D. S. Madgwick, P.
Norberg, J. A. Peacock, I. K. Baldry, C. M. Baugh, J.
Bland-Hawthorn, T. Bridges, R. Cannon, M. Colless, C. Collins, W.
Couch, G. Dalton, R. D. Propris, S. P. Driver, S.P., G. Efstathiou,
R. S. Ellis, C.S. Frenk, K. Glazebrook, C. Jackson, B. Jones, I.
Lewis, S. Lumsden, W. Percival, B. A. Peterson, W. Sutherland, and
K. Taylor: Mon. Not. Roy. Astr. Soc.
\textbf{346} (2003) 78.\\
7) B. Jain and A. Taylor: Phys. Rev. Lett.
\textbf{91} (2003) 141302.\\
8) A. S. Al-Rawaf and M. O. Taha: Gen. Relativ. Gravit.
\textbf{28} (1996) 935.\\
9) R. R. Caldwell, R. Dave, and P. J. Steinhardt: Phys. Rev. Lett.
\textbf{80} (1998) 1582.\\
10) P. J. Steinhardt, L. Wang, and I. Zlatev: Phys. Rev. Lett.
\textbf{59} (1999) 123504.\\
11) I. Zlatev, L. Wang, and P. J. Steinhardt: Phys. Rev. Lett.
\textbf{82} (1999) 896.\\
12) L. A. U. Lopez and T. Matos: Phys. Rev. D \textbf{62} (2000) 081302.\\
13) O. Bertolami and P. J. Martins: Phys. Rev. D \textbf{61} (2000) 064007.\\
14) N. Benerjee and D. Pavon: Phys. Rev. D \textbf{63} (2001) 043504.\\
15) B. K. Sahoo and L. P. Singh: Mod. Phys. Lett. A \textbf{18} (2003) 2725.\\
16) W. Chakraborty and U. Debnath: Int. J. Theor. Phys. \textbf{48}(2009)232.\\
17) C. H. Brans and R. H. Dicke: Phys. Rev. \textbf{124}(1961)925.\\
18) S. Weinberg: \emph{Gravitation and Cosmology} (Wiley, 1972).\\
19) P. A. M. Dirac: Proc. R. Soc. Lond. A \textbf{165} (1938) 199.\\
20) B. Bertotti, L. Iess, and P. Tortora: Nature
\textbf{425} (2003) 374.\\
21) A. D. Felice, G. Mangano, P. D. Serpico, and M. Trodden: Phys.
Rev. D \textbf{74} (2006) 103005.\\
22) S. K. Rama and S. Gosh: Phys. Lett. B \textbf{383} (1996) 32;
S. K. Rama: Phys. Lett. B \textbf{373} (1996) 282.\\
23) C. Romero and A. Barros: Phys. Lett. A \textbf{173} (1993) 243.\\
24) N. Benerjee and S. Sen: Phys. Rev. D \textbf{56} (1997) 1334.\\
25) J. P. Singh and P. S. Baghel: Elect. J. Theor. Phys.
\textbf{6} (2009) 85.\\
26) M. K. Verma, M. Zeyauddin, and S. Ram: Rom. J. Phys.
\textbf{56} (2011) 616.\\
27) H. K. Eriksen, F. K. Hansen, A. J. Banday,
K. M. Gorski, and P. B. Lilje: Astrophys. J. \textbf{605} (2004) 14.\\
28) T. R. Jaffe, A. J. Banday, H. K. Eriksen, K. M. Gorski, and
F. K. Hansen: Astrophys. J. \textbf{629} (2005) L1.\\
29) E. G. King and P. Coles: Class. Quantum Grav.
\textbf{24} (2007) 2061.\\
30) T. Singh and L. N. Rai: Gen. Relativ. Gravit.
\textbf{15} (1983) 875.\\
31) M. Sharif and S. Waheed: Eur. Phys. J. C \textbf{72} (2012) 1876.\\
32) M. S. Bermann: Nuovo Cimento B \textbf{74} (1983) 192.\\
33) J. Khoury and A. Weltman: Phys. Rev. D \textbf{69} (2004) 044026.\\
34) R. de Ritis, A. A. Marino, C. Rubano, and P. Scudellaro: Phys.
Rev. D \textbf{62} (2000) 043506.\\
35) S. Sen and A. A. Sen: Phys. Rev. D \textbf{63} (2001) 124006.\\
36) V. Faraoni: Phys. Rev. D \textbf{62} (2000) 023504.\\
37) S. Sen and T. R. Seshadri: Int. J. Mod. Phys. D \textbf{12} (2003) 445.\\
38) M. Sharif and M. Zubair: Astrophys. Space Sci.
\textbf{330} (2010) 399.\\
39) C. B. Collins, E. N. Glass, and D. A. Wilkinson: Gen. Relativ.
Gravit. \textbf{12} (1980) 805.\\
40) K. S. Throne: Astrophys. J. \textbf{148} (1967) 51.\\
41) J. Kristian and R. K. Sachs: Astrophys. J.
\textbf{143} (1966) 379.\\
42) C. B. Collins: Phys. Lett. A \textbf{60} (1977) 397.\\
43) J. Khoury and A. Weltman: Phys. Rev. Lett. \textbf{93} (2004)
171104.\\
44) J. Khoury and A. Weltman: Phys. Rev. D \textbf{69} (2004)
044026.\\
45) S. Tsujikawa, K. Uddin, S. Mizuno, R. Tavakol, and J. Yokoyama:
Phys. Rev. D \textbf{77} (2008) 103009.\\
46) S. Tsujikawa, T. Tamaki, and R. Tavakol: JCAP \textbf{0905} (2009) 020.\\
47) D. F. Mota and H. A. Winther: Astrophys. J. \textbf{733} (2011) 7.\\
48) R. R. Caldwell: Phys. Lett. B \textbf{545} (2002) 23.\\
49) A. A. Sen, S. Sen, and S. Sethi: Phys. Rev. D
\textbf{63} (2001) 107501.\\
50) A. F. Bahrehbakhsh, M. Farhoudi, and H. Shojaie: Gen. Relativ.
Grav. \textbf{43} (2011) 847.\\
51) J. P. D. Leon: Class. Quantum Grav.
\textbf{27} (2010) 095002.\\
52) E. Gaztanaga and J. A. Lobo: Astrophys. J.
\textbf{548} (2001) 47.\\
53) L. Perivolaropoulos: Phys. Rev. D \textbf{67} (2003) 123516.\\
54) L. Perivolaropoulos: Phys. Rev. D \textbf{81} (2010) 047501.

\end{document}